\documentclass[prd,twocolumn,superscriptaddress,preprintnumbers,nofootinbib]{revtex4-2}
\usepackage{graphicx}
\usepackage{epsfig}
\usepackage{bm} 
\usepackage{latexsym,amssymb,amsmath,amsfonts,amssymb,txfonts,wasysym,float}
\usepackage{mathrsfs}
\usepackage{color}

\usepackage{enumitem}

\newcommand{\postscript}[2]{\setlength{\epsfxsize}{#2\hsize}
   \centerline{\epsfbox{#1}}}

\usepackage[usenames,dvipsnames]{xcolor}
\definecolor{orange}{cmyk}{0,0.5,1,0}
\definecolor{rossoCP3}{cmyk}{0,.88,.77,.40}
\definecolor{graa}{rgb}{0.8,0.8,0.8}
\definecolor{blaa}{rgb}{0.2,0.2,0.6}

\begin{document}

\title{\color{rossoCP3} Sombrero Galaxy as an Accelerator of Ultrahigh
  Energy
  Cosmic Ray Nuclei}

\author{\bf Luis A. Anchordoqui}\thanks{\tt luis.anchordoqui@gmail.com}

\affiliation{Department of Physics and Astronomy, Lehman College, City University of
  New York, NY 10468, USA
}

\affiliation{Department of Physics,
 Graduate Center,  City University of
  New York,  NY 10016, USA
}

\affiliation{Department of Astrophysics,
 American Museum of Natural History, NY
 10024, USA
}

\author{\bf Karem Pe\~nal\'o Castillo}\thanks{\tt karem.penalo@gmail.com}

\affiliation{Department of Physics and Astronomy, Lehman College, City University of
  New York, NY 10468, USA
}

\begin{abstract}
  \noindent Motivated by a recent proposal that points to the Sombrero
  galaxy as a source of the highest energy cosmic rays, we investigate
  the feasibility of accelerating both light and heavy nuclei near the
  supermassive black hole located at the center of this nearly edge-on
  galaxy that is just beyond naked-eye visibility. We show that cosmic
  ray nuclei concentrated in the immediate vicinity of the
  supermassive black hole could be efficiently accelerated up to the
  maximum observed energies without suffering catastrophic
  spallations. Armed with our findings we stand against conventional
  wisdom and conjecture that accelerators of the highest energy cosmic
  rays must anti-correlate with the (electromagnetic) source power.
\end{abstract}
\date{March 2025}
\maketitle

\section{General Idea}

 There exists lore advocating that the accelerators of
 ultrahigh energy cosmic rays (UHECRs) must correlate with the
 (electromagnetic) source
 power. Certainly, to account for the observed UHECR flux these sources
 must be injecting energy into the cosmos at a rate of $\sim 10^{37.3}~{\rm TeV/s/Mpc^3}$~\cite{Waxman:1995dg}. This 
 works out to: {\it (i)}~$\sim  10^{39.3}~{\rm TeV/s}$ per galaxy,
{\it (ii)}~$\sim 10^{42.3}~{\rm TeV/s}$ per cluster of galaxies,
{\it (iii)}~$\sim  10^{44}~{\rm TeV/s}$ per active galaxy, or
{\it (iv)}~$\sim 10^{52}~{\rm TeV}$ per cosmological gamma-ray burst~\cite{Gaisser:1997aw}. The coincidence between these numbers and the observed output in electromagnetic energy of these sources explains why they have emerged as the leading candidates for cosmic ray accelerators~\cite{Anchordoqui:2018qom}. On
 that account, it is somewhat
 unexpected that the back-tracked directions from two of the highest
 energy cosmic rays (those with the smallest localization
 uncertainties on the source direction) do not 
align with any compelling powerful
accelerator~\cite{Unger:2023hnu,Unger:2025xkk}. This null result has
been exploited to speculate that a fraction of the highest energy
events must originate from transient
sources~\cite{Unger:2025xkk}.

In this paper we argue in favor of a totally distinct viewpoint. Namely, we propose that UHECR sources are anti-correlated
with the (electromagnetic) source power; more specifically, with
emission of photons at infrared (IR) and far infrared (FIR) frequencies. Our proposal builds upon an idea introduced in
the late nineties, according to which UHECRs could be accelerated near
the event horizons of spinning supermassive black holes~\cite{Boldt:1999ge}. UHECR acceleration proceeds via
the Blandford-Znajek process, which wires spinning black
holes as Faraday unipolar dynamos~\cite{Blandford:1977ds}. The
essence of our proposal, however, does not derive only from the
acceleration mechanism, but also from energy loss in the source
environment. It is not hard to imagine that if UHECR sources were not powerful galaxies to our eyes, then they would have an associated low photon density inside the accelerator, allowing for some UHECRs to escape the source
without significant energy losses. Furthermore, throughout we subscribe to the so-called ``UFA model'', wherein
photo-disintegration of ultrahigh energy nuclei in the region
surrounding UHECR accelerators could account for the observed spectrum and inferred composition at Earth~\cite{Unger:2015laa}.

Very recently, it was pointed out that the Sombrero galaxy could be a
source of UHECRs~\cite{He:2024vnm}. Indeed, the back-tracked directions from multiple UHECRs
were shown to align with the location of Sombrero. The 
multiplet,
which contains one of the highest energy events analyzed
in~\cite{Unger:2025xkk}, was
claimed to have a global significance of $3.3\sigma$. Of particular
interest for our proposal, 
Sombrero is classified as a low-luminosity active galactic
nucleus~\cite{Goold}, with a supermassive black hole located at its center, accreting at less than 1\% of its Eddington limit~\cite{Kormendy:1988,
  Kormendy:1996,Jardel:2011dh}. In this paper, we reexamine whether it is feasible for Sombrero to emit UHECR nuclei. 

The layout is as follows: In Sec.~\ref{sec:2} we provide a short
overview of estimated deflections of the highest energy cosmic rays
by the Galactic magnetic field, and how Sombrero became a plausible UHECR source. In Sec.~\ref{sec:3} we describe the current
knowledge of the main properties of Sombrero based on the large
collection of data from different instruments. In Sec.~\ref{sec:4} we
derive the 
characteristic timescales for acceleration and energy loss, and
by equating these relations we show that Sombrero can indeed be a compelling
source of UHECR nuclei. Lastly, in Sec.~\ref{sec:5} we discuss the possible emission of $\gamma$-rays produced by curvature radiation of UHECRs. The paper wraps up in Sec.~\ref{sec:6} with some conclusions.

Before proceeding, we pause to briefly examine yet another
alternative. It has been generally thought that $^{56}$Fe is a
significant end product of stellar evolution, and higher mass nuclei
are rare in the cosmic radiation. Strictly speaking, the 
abundances of  supermiddle-weight ($60 \leq A < 100$) and superheavy-weight ($A >
100$) nuclei are respectively about three and five orders of magnitude lower than that of the iron
group~\cite{Burbidge:1957vc}. Since the late nineties, however, it has already been evident that the UHECR horizon, which is the largest distance that cosmic rays can propagate at a given energy, would be enlarged if the highest energy cosmic rays are superheavy
nuclei~\cite{Anchordoqui:1999aj,Anchordoqui:2000sk}. Indeed, if this were the case, the larger number of available sources
could compensate for the lower abundance of this type of
nuclei. Recently, this idea has been revitalized by the possibility
of UHECR production in neutron star
mergers~\cite{Farrar:2024zsm,Zhang:2024sjp}. Such an exciting possibility then combines the
 alternative of a larger horizon with the practicality of transient
 sources. 

\section{UHECR Deflections}
\label{sec:2}
 
Galactic and extragalactic magnetic fields significantly deflect UHECRs
as they journey to Earth. This deflection obscures the true source
locations, preventing us from simply matching the observed UHECR
arrival directions with catalogs of potential astronomical sources. To
overcome this obstacle, one can try to reconstruct the UHECR
trajectories backwards using existing magnetic field models. This
back-tracking method holds particular promise for the highest-energy
events, where the expected deflections are comparatively smaller.

In the studies of~\cite{Unger:2023hnu,Unger:2025xkk} the arrival directions of the five highest energy
cosmic-ray events detected by the Telescope Array (TA) and the Pierre Auger
Observatory (PAO) were backtracked through the Galactic magnetic
field~\cite{Jansson:2012pc,Unger:2023lob} under the assumption that all UHECRs are iron nuclei. The key
parameters characterizing the UHECR events are summarized in
Table~\ref{tabla}. The arrival directions of the two events with the smallest
localization uncertainty (PAO070114 and TA210527) are far from the Galactic plane, and
consequently their paths avoid regions of high magnetic field
strength. However, the trajectories of the other three events pass
close to the Galactic plane, where the
realization-to-realization variance of the random Galactic magnetic
field is large. For the two UHECR events with the smallest
localization uncertainties, neither of their
back-tracked directions align with any compelling powerful UHECR
accelerator~\cite{Unger:2025xkk}.

It is worth commenting on some caveats of the analysis in~\cite{Unger:2025xkk}: 
\begin{itemize}[noitemsep,topsep=0pt]
\item The study of~\cite{Unger:2025xkk} assumes that cosmic rays at the high energy end
  of the spectrum are $^{56}$Fe nuclei. We note that the UHECR horizon would be even smaller for lighter nuclei, but there could be a lucky coincidence with a nearby
object. Obviously, the deflections would be then somewhere in between
those of
$^{56}$Fe given in Table~\ref{tabla} and the UHECR arrival directions.
\item The Galactic magnetic field models~\cite{Jansson:2012pc,Unger:2023lob} adopted in the analysis of~\cite{Unger:2025xkk} do not
  bracket the UHECR deflection uncertainty. Other models of the
  Galactic magnetic field (e.g.~\cite{Korochkin:2024yit}) could lead to larger localization uncertainties~\cite{Korochkin:2025ugg}. 
\item Alternatively, a larger localization uncertainty than the one
  given in Table~\ref{tabla} would be obtained if there were a significant
  contribution to the UHECR deflection from the extragalactic magnetic field.
\item Yet another way of enlarging the localization uncertainty would
  be to consider that UHECRs are superheavy nuclei, for which the
  horizon distance of their sources  is significantly larger than for iron nuclei. 
\end{itemize}

A search for UHECR multiplets in Auger data~\cite {PierreAuger:2022axr} led to a spatial
association between a multiplet of $25.7^{+6.2}_{-7.0}$ cosmic rays
(of $E> 40~{\rm EeV}$) and the Sombrero galaxy, with a local (global)
significance of $4.5~\sigma$ ($3.3~\sigma$)~\cite{He:2024vnm}. The
multiplet contains the event with the smallest localization
uncertainty in Table~\ref{tabla}. This intriguing finding motivates
Sombrero as an UHECR accelerator and, leaving aside the caveats
itemized above, it aligns well with our conjecture.

\begin{table}
  \caption{Key parameters (energy with statistical uncertainty and
    arrival direction in Galactic coordinates) of the UHECR events used in the analysis of~\cite{Unger:2025xkk}. The quoted energies are at the Auger
    energy scale and the errors are statistical. The last column
    stands for the percentage localization uncertainty on the source
    direction of $4 \pi$. The numerical value of
the localization of TA\-210527 is somewhat larger
in~\cite{Unger:2023hnu} (6.6\% instead of 4.7\%) because it includes the
uncertainty in the TA energy scale.
\label{tabla}}
 \begin{tabular}{lcccc|c}
\hline
\hline
   UHECR id & $E$ & $\sigma_\text{stat.}$   & \multicolumn{1}{c}{$\ell$}
   & \multicolumn{1}{c|}{$b$} & ~~$\Omega_\text{loc}$ / $(4\pi)$~~ \\
       & (EeV) & ~~(EeV)~~  & \multicolumn{1}{c}{~~(degree)~~} & \multicolumn{1}{c|}{~~(degree)~~} & --  \\ \hline
PAO191110~~ &166 &13  &269.1 & \phantom{$4$}$-6.8$ & 7.1\% \\
PAO070114 &165 &13  & 303.0 & \phantom{$-$}$41.7$ & 2.4\% \\
PAO141021 &155 &12 & 247.6 & $-16.2$ & 6.3\% \\
PAO200611 &155 &12  & 258.3 & $-16.9$& 6.6\% \\
TA210527 & 154 & 18  &  \phantom{$2$}36.2 & \phantom{$-$}$30.9$& 4.7\% \\
   \hline
   \hline
 \end{tabular}
\end{table}

\section{Key Features of  Sombrero Galaxy}
\label{sec:3}

\begin{figure}[tbh]
  \postscript{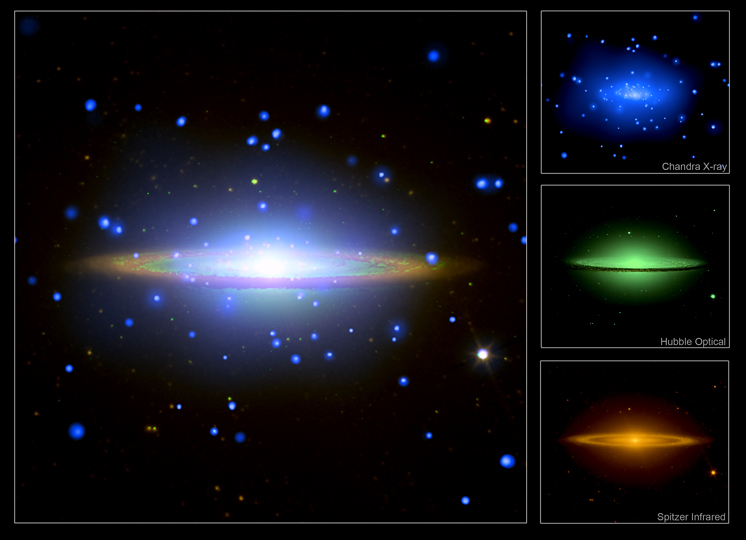}{0.95}
  \caption{The Sombrero galaxy as seen by Chandra, Hubble, and Spitzer
    observations. Chandra's X-ray image (blue) shows hot gas in the
    galaxy and point sources that are a mixture of objects within the
    galaxy and quasars in the background. Hubble's optical image
    (green) reveals the bulge of starlight partially blocked by a rim
    of dust, which glows brightly in Spitzer's infrared view
    (orange). The image scale is 8.4~arcmin per side. Credit: X-ray, NASA/UMass/Q.D.Wang et al.; Optical, NASA/STScI/AURA/Hubble Heritage; Infrared, NASA/JPL-Caltech/Univ. AZ/R.Kennicutt/SINGS Team. \label{fig:1}}
\end{figure}

The Sombrero Galaxy (a.k.a. M104 and NGC 4594) has a bright nucleus,
an unusually large central bulge, and a prominent dust lane in its
outer disk, which from Earth is viewed almost edge-on. The dark dust
lane and the bulge give it the appearance of a broad-brimmed Mexican
hat floating in space; see Fig.~\ref{fig:1}. This
galactic hat, located about $9~{\rm Mpc}$ away~\cite{McQuinn}, has total stellar mass $\sim 10^{11.3} M_\odot$~\cite{Tempel:2006qu}.

Spectroscopic data from both the Canada-France-Hawaii Telescope and the Hubble Space Telescope (HST)
indicate that the speed of revolution of the stars within the center
of the galaxy could not be maintained unless its nucleus hosts a
supermassive black hole, with a mass $M_\bullet \sim 10^9 M_\odot$ ~\cite{Kormendy:1988, Kormendy:1996,Jardel:2011dh}. However, the
bolometric nuclear luminosity is about 200 times lower than expected
if mass is accreted on the supermassive black
hole at the rate predicted by the spherical and
adiabatic Bondi accretion theory~\cite{Bondi:1952ni} and with the high radiative
efficiency of a standard accretion
disk. To explain the very sub-Eddington luminosity of Sombrero’s
nucleus the Bondi mass accretion rate is estimated to be 
\begin{equation}
  0.008 \alt \dot M_{\rm Bondi}~{\rm yr}/M_\odot \alt 0.067 \,,
\end{equation}
with a best fit value of $\dot M_{\rm Bondi} \simeq 0.013~M_\odot/{\rm yr}$~\cite{Pellegrini:2003pu,Li:2010uk}. This discrepancy suggests that a large
fraction of the accretion luminosity could be stored in the form of
jet power. Indeed Very-Long-Baseline-Interferometry images toward the nucleus are consistent with the
presence of kpc-scale nuclear radio jets~\cite{Gallimore:2006dp,Hada:2013tqa}. Furthermore,  bipolar radio lobes have been detected on the 10-kpc scale~~\cite{Yang}.

Radio polarization and sub-mm observations lead to an estimate of the
average large-scale magnetic field strength $\overline{B}_{\rm{bulge}}
\sim 4~\mu{\rm G}$~\cite{Krause:2005bt}. The estimation has been done
taking an 
average over
the entire galaxy excluding the nucleus~\cite{Yang}.  Deep, high-resolution HST imaging shows that the
halo of the Sombrero galaxy is strikingly metal-rich, with its metallicity distribution function peaking at
near-solar metallicity~\cite{Cohen:2020}. This is the highest peak
metallicity found in halos of nearby galaxies to
date.

The FIR to sub-mm flux densities of Sombrero are rather
weak compared to other galaxies. The average IR luminosity of Sombrero as given in~\cite{Young:1989} (using the
IRAS survey~\cite{IRAS}) but corrected for a distance of 9~Mpc is
\begin{equation}
  \log_{10}(\overline{L_{\rm IR}^{\rm tot}}/L_\odot) = 9.3 \, ,
\end{equation}  
whereas  the FIR luminosity of
its nucleus estimated in~\cite{Sutter:2022} (using Herschel, ALMA, and MUSE observations) but corrected for a distance of 9~Mpc is
\begin{equation}
  \log_{10}(\overline{L_{\rm FIR}^{\rm nuc}}/L_\odot) = 6.7 \, .
\end{equation}  
The ratio of these luminosities is in good agreement with Spitzer and
JCMT observations of the flux density in the nucleus, inner-disk,
ring, and bulge~\cite{Bendo:2006pc}.

\section{UHECR Emission from Sombrero Galaxy}
\label{sec:4}

It is well-established that a rotating black hole can radiate away its
available reducible
energy~\cite{Penrose:1969pc,Penrose:1971uk}.\footnote{In black hole
  physics, the total mass-energy can be divided into reducible and
  irreducible components~\cite{Christodoulou:1970wf}. The irreducible mass
  represents the minimum mass a black hole must have, and it is
  related to the area of the black hole's event horizon. The reducible
  mass, also known as rotational (or electrostatic) energy, is the
  energy that can potentially be extracted from the black hole.} In
particular, Blandford and Znajek proposed a compelling process,
analogous to Faraday's classical (unipolar induction) dynamo, by which
the spin energy of a black hole might be extracted by force-free
electromagnetic fields~\cite{Blandford:1977ds}. The dynamo's electromotive force (emf) 
is produced by the black hole induced rotation of magnetic field lines
in the inermost accretion disk.

\begin{figure*}[htb!]
  \begin{minipage}[t]{0.49\textwidth}
    \postscript{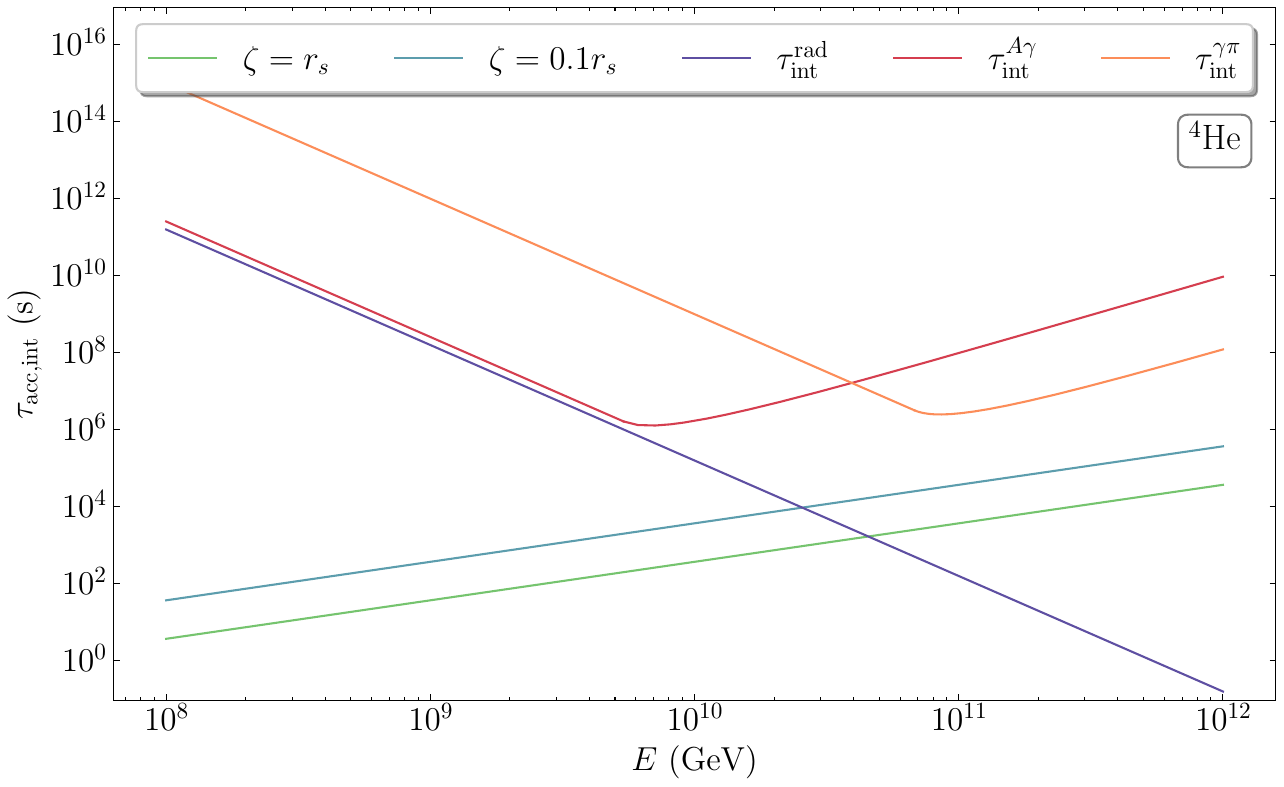}{0.99}
  \end{minipage}
\begin{minipage}[t]{0.49\textwidth}
    \postscript{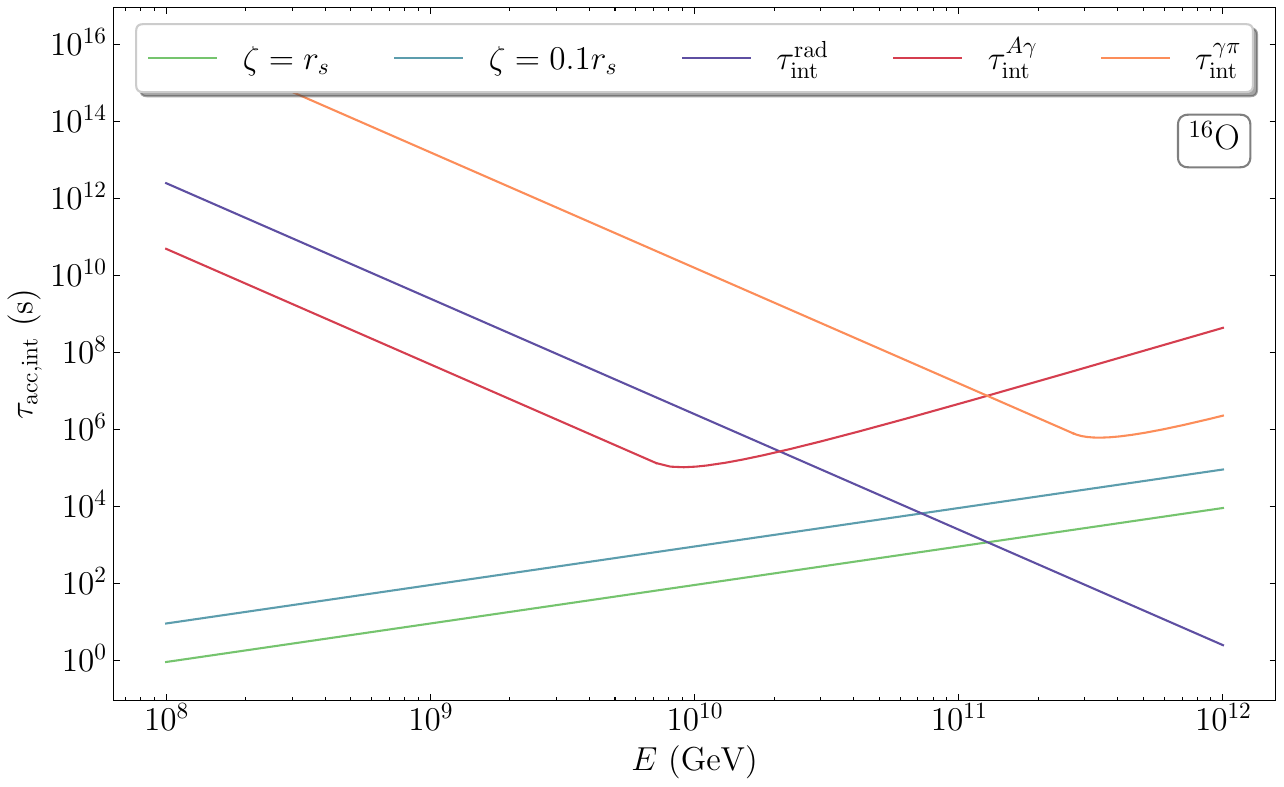}{0.99}
  \end{minipage}
   \begin{minipage}[t]{0.49\textwidth}
    \postscript{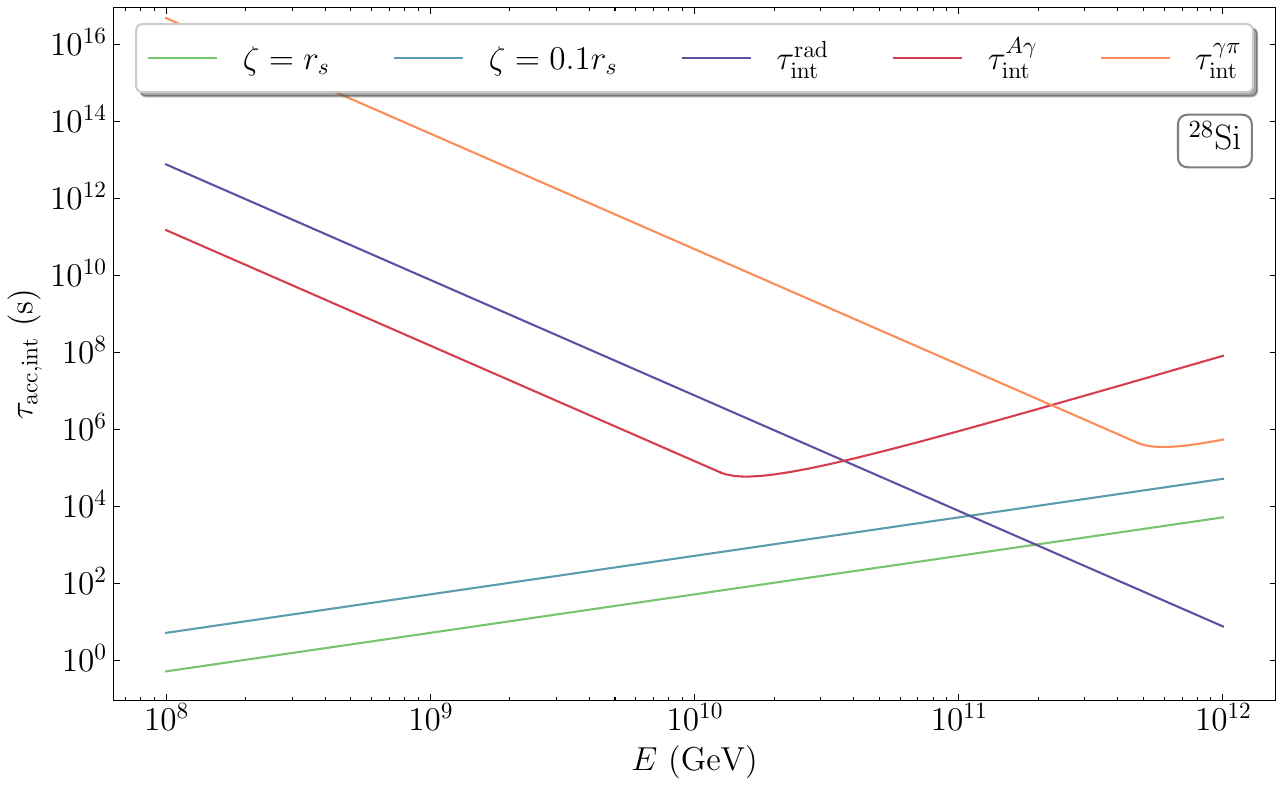}{0.99}
  \end{minipage}
\begin{minipage}[t]{0.49\textwidth}
    \postscript{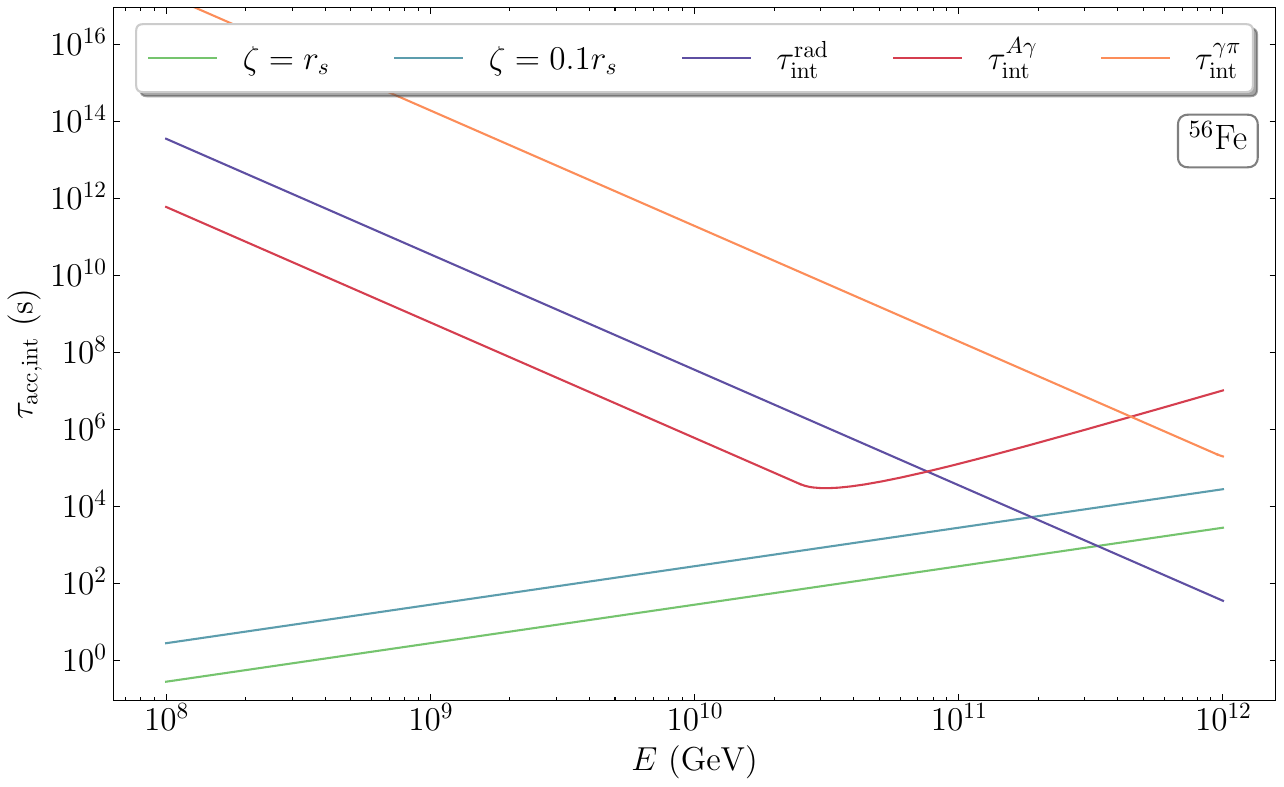}{0.99}
  \end{minipage}
  \caption{Comparison of acceleration and interaction timescales. The
    straight lines with positive slope correspond to acceleration
    timescales of different gap hight $\zeta$. The straight line with negative
    slope indicates the timescale due to curvature radiation. The $V$-shaped curves indicate the interaction timescales of photonuclear processes.  The different panels correspond to helium (upper left), oxygen (upper right), silicon (bottom left) and iron (bottom right). \label{fig:2}}
  \end{figure*}

More concretely, spinning black holes warp spacetime around them, creating a region
called the ergosphere where spacetime rotates with the black hole. A
magnetic field, externally supplied, threads this ergosphere. The
rotating spacetime drags the magnetic field lines along with it. This
dragging action twists and stretches the magnetic field lines,
inducing an electric field and thus an emf ($V$) that is estimated to be
\begin{equation}
   V \sim a B\,,
\label{V1}
 \end{equation}
where $B$ is the poloidal magnetic field and $a$ is the black hole's specific
 angular momentum, with $a \leq
 M_\bullet$~\cite{Znajek}.\footnote{Following~\cite{Blandford:1977ds},
   $M_\bullet$ and $a$ are measured in units of $c^2/G$ and $c$, respectively, so that both quantities have dimensions of
length. Then $M_\bullet$ and $ac/M_\bullet^2$ respectively provide a length scale and
a characteristic angular frequency for the spinning black hole, which
has angular momentum $J = a M_\bullet$.} In
 astrophysical units (\ref{V1}) can be rewritten as
\begin{equation}
V \sim 9 \times 10^{11} \ \left(\frac{a}{M_\bullet}\right) \
\left(\frac{M_\bullet}{10^9 M_\odot}\right)  \ \left(\frac{B}{10^4~G}\right)~{\rm GV} \, .
\label{V2}
\end{equation}
Around the event horizon the energy density of the $B$ field is
expected to be in equipartition with the rest mass energy density of
accreting matter~\cite{Krolik:1999yc}. In connection to the advection dominated
accretion flow model discussed in~\cite{DiMatteo:1998bf}, this is to be
identified with the regime where the gas pressure is half the
total. Bearing this in mind, the accretion rate is estimated to be~\cite{Boldt:2000dx}
\begin{equation}
\frac{dM_\bullet}{dt} \sim \frac{9}{16} \ \left(\frac{B}{10^4~{\rm G}}\right)^2
  \left(\frac{M_\bullet}{10^9 M_\odot} \right)^2~{\rm M_\odot/yr} \, .
\label{mdot}
\end{equation}   
Now, assuming that the black hole accretes at the Bondi rate
and substituting (\ref{mdot}) into (\ref{V2}), we consider the case of a rapidly spinning black hole with $(a/M_\bullet)^2$ not much
smaller than unity, and express the maximum emf directly in terms of the accretion rate
\begin{equation} 
  V =  1.2 \times 10^{12} \  \left(\frac{\dot M_{\rm Bondi}}{M_\odot/{\rm yr}}\right)^{1/2}~{\rm GV} \, .
\label{V3}
\end{equation}  

The charge density around the horizons of accreting black holes could
be high.\footnote{In essence, for sufficiently low accretion
  rates, the flow becomes radiatively inefficient and the 
  temperature of the hot gas near the horizon could exceed
  $m_e$~\cite{Narayan:1994is}. Bremsstrahlung cooling would thus give rise to  emission of MeV
  photons, which would annihilate in the black hole magnetosphere leading to
  the injection of charges on the magnetic field
  lines~\cite{Levinson:2010fc}. When the charge density in the
  magnetosphere is sufficiently small, regions of unscreened electric
  fields (i.e., gaps) form. For a 
  sufficiently high annihilation rate, however, the resultant charge density can
  exceed the Goldreich-Julian value, screening off the electric field
  component~\cite{Neronov:2009zz}. For further details see e.g.~\cite{Katsoulakos:2017byr}.}  If this were the case, a significant fraction of emf given
in (\ref{V3}) would be screened and not available for UHECR
acceleration. To consider this effect, following~\cite{Rieger:2011ch}, we introduce
the effective potential,
\begin{equation}
  V_{\rm eff} \sim V \ \left(\frac{\zeta}{r_s} \right)^2 \,,
\end{equation}
which explicitly takes into account the gap height $\zeta$ as the
available length scale for acceleration, where $r_s = 2GM_\bullet$ is
the Schwarzschild radius. All in all, the characteristic rate of energy gain is found to be
\begin{equation}
\left. \frac{dE}{dt} \right|_{\rm acc} = Z e \ V_{\rm eff} \ \frac{c}{\zeta} \,,
\end{equation}
and the acceleration timescale is given by
\begin{equation}
\tau_{\rm acc}  = E \left[\left. \frac{dE}{dt}
  \right|_{\rm acc} \right]^{-1} = \frac{E}{Z e V_{\rm eff}}  \,  \frac{\zeta}{c} \, .
\label{muddler}
\end{equation}

Next, in line with our stated plan, we turn to investigate the energy
loss. Inside the potential drop, the UHECR nuclei follow the curved magnetic field lines and hence they emit curvature-radiation photons. The energy loss rate or total power radiated away by a single cosmic ray is~\cite{Ochelkov} 
\begin{equation}
\left. \frac{d E}{dt} \right|_{\rm loss}^{\rm rad} = \frac{2}{3} \frac{Z^2 e^2 c}{r_c^2}
\gamma^4 \,,
\end{equation}
where $r_c \sim r_s$ is the curvature radius of the magnetic field lines, and
$\gamma$ is the Lorentz factor of the radiating particles. 
The characteristic interaction timescale is then estimated to be
\begin{equation}
\tau_{\rm int}^{\rm rad}  =  \gamma A m_p  \left[\left. \frac{d
      E}{dt} \right|_{\rm loss}^{\rm rad} \right]^{-1}   \, ,
\end{equation}
where $A$ is the baryonic number of the nucleus and $m_p$ is the
proton mass. 

A comparison between the expected acceleration time and loss time
allows us to give an estimate of the maximum energy reachable by
UHECR nuclei. Indeed the maximum energy of the cosmic ray is given by
the value at which energy lost and gained  are equal. In Fig.~\ref{fig:2} we show the characteristic timescales for UHECR acceleration and energy dissipation in the black hole
dynamo. As one can check by inspection of this
figure, heavy ($^{28}$Si and $^{56}$Fe) nuclei can be accelerated by
the supermassive black hole of Sombrero up to well above the maximum energies
observed on Earth. However, this is not the case for lighter
nuclei. For example, for $^{4}$He the maximum energy UHECR can reach
is somewhat below $10^{11}~{\rm GeV}$, whereas for $^{16}$O the
maximum energy is just above  $10^{11}~{\rm GeV}$.

It was shown elsewhere~\cite{Moncada:2017hvq} that nucleus
photodisintegration in the vicinity of the black hole could suppress the emission of UHECR nuclei. We now shift our focus to study this effect. In keeping with the UFA model, we approximate the isotropic photon background as a broken power-law,
\begin{equation}
      n(\varepsilon) = n_0
        \begin{cases}
           (\varepsilon/\varepsilon_0)^{\, \alpha} & \quad \varepsilon < \varepsilon_0 \\
           (\varepsilon/\varepsilon_0)^{\, \beta} & \quad \text{otherwise} 
        \end{cases} \,,
\label{app:eq:photonfield}
\end{equation}
where $\varepsilon$ is the
photon energy and the maximum of the density is at an energy of
$\varepsilon_0 \sim 0.01~{\rm eV}$. The broken power-law spectrum allows a complete
analytic treatment of photonuclear interactions, and the global result
does not depend on the exact shape of the photon spectrum.  Namely,
the interaction times are comparable if the photon density is assumed
to follow a (modified) black body spectrum~\cite{Unger:2015laa}. To simplify the analytical treatment we
take $\alpha = 2$ and $ \beta = -4$. This choice of parameters properly approximates the region of the spectrum relevant to nucleus photodisintegration. 

 We normalize the spectrum to the FIR luminosity of Sombrero's
 nucleus. More precisely, we adopt an average of the FIR luminosity over the nucleus' spatial volume taken to be a sphere of radius $r_{\rm nuc} \sim 50 \, r_s$~\cite{Hada:2013tqa},
\begin{eqnarray}
\overline{L^{\rm nuc}_{{\rm FIR}}} \! & = & \! 4\pi r_{\rm nuc}^2c\int
n(\varepsilon)\,\varepsilon\,d\varepsilon \nonumber \\
& = & 4\pi\,r_{\rm nuc}^2\,c\,n_0\left\{\int_{\varepsilon_{\rm
      min}}^{\varepsilon_0} \! \! \! \!\!
  (\varepsilon/\varepsilon_0)^\alpha\,\varepsilon\,d\varepsilon+ \!\! \int_{\varepsilon_0}^\infty
  \! \! \! \! (\varepsilon/\varepsilon_0)^{\beta}\,\varepsilon\,d\varepsilon\right\}\!\!,
\end{eqnarray}
where $\varepsilon_{\rm min} = 10^{-4}~{\rm eV}$.\footnote{The characteristic energy for the transition between
microwave and IR photons is of the order of {\rm meV}. To remain
conservative herein we follow~\cite{Moncada:2017hvq} and adopt $\varepsilon_{\rm min} \sim 10^{-4}~{\rm
  eV}$.} Putting all this
together, the normalization factor is given by
\begin{equation}
n_0=\frac{\overline{L_{\rm FIR}^{\rm nuc}}}{4\pi\,r_{\rm nuc}^2\,c} 
  \left\{\frac{\varepsilon_0^{-\alpha}}{(\alpha+2)}\left[\varepsilon_0^{\alpha+2}-\varepsilon_{\rm
        min}^{\alpha+2}\right] -
    \frac{\varepsilon_0^2}{(\beta+2)} \right\}^{-1} \, ,
  \label{n0}
\end{equation}
or equivalently $n_0 =  10^{18.3}~{\rm MeV^{-1} cm^{-3}}$. 

The interaction time for a highly relativistic nucleus propagating
through an isotropic photon background with energy $\varepsilon$ and spectrum
$n(\varepsilon)$, normalized so that the total number of photons in a
box is $\int n ( \varepsilon ) d \varepsilon$, is given by
\begin{equation}
 \frac{1}{\tau_\mathrm{int}^{A\gamma}} = \frac{c}{2} \,\int_0^\infty
                 d\varepsilon \,\frac{n(\varepsilon)}{\gamma^2 \varepsilon^2}\, \int_0^{2\gamma\varepsilon}
                 d\varepsilon^\prime \, \varepsilon^\prime\, \sigma(\varepsilon^\prime),
\label{app:eq:interaction}
\end{equation}
where $\sigma(\varepsilon^\prime)$ is the photonuclear interaction
cross-section of a nucleus and a photon of energy
$\varepsilon'$, in the rest frame of the nucleus~\cite{Stecker:1969fw}.

We have found that for the considerations in the present work, the cross section can be safely
approximated by the single pole of the narrow-width approximation,
\begin{equation}
\sigma (\varepsilon') = \pi\,\,\sigma_{\rm res}\,\,  \frac{\Gamma_{\rm
  res}}{2} \,\,
\delta(\varepsilon' - \varepsilon'_{\rm res})\, ,
\label{sigma}
\end{equation}
where $\sigma_{\rm res}$ is the resonance peak, $\Gamma_{\rm res}$ its
width, and $\varepsilon'_{\rm res}$ the pole in the rest frame of the
nucleus.  The factor of $1/2$ is introduced to match the integral
(i.e. total cross section) of the Breit-Wigner and the delta
function~\cite{Anchordoqui:2006pe}.

The mean interaction time  can now be readily
obtained  substituting Eq.~(\ref{sigma}) into Eq.~(\ref{app:eq:interaction}),
\begin{eqnarray}
  \frac{1}{\tau_{\rm int}^{A\gamma} (E)} & \approx & \frac{c\, \pi\,
    \sigma_{\rm res}
    \,\varepsilon'_{\rm res}\,
\Gamma_{\rm res}}{4\,
    \gamma^2}
  \int_0^\infty \frac{d \varepsilon}{\varepsilon^2}\,\,\, n(\varepsilon) \,\,\,
  \Theta (2 \gamma \varepsilon - \varepsilon'_{\rm res}) \nonumber \\
  & = & \frac{c \, \pi \, \sigma_{\rm res} \,\varepsilon'_{\rm res}\,
    \Gamma_{\rm res}}{4 \gamma^2}
  \int_{\epsilon'_{\rm res}/2 \gamma}^\infty \frac{d\varepsilon}{\varepsilon^2}\,\,
  n (\varepsilon)  \, .
 \label{A1}
\end{eqnarray}

Substituting (\ref{app:eq:photonfield}) into (\ref{A1}) we finally find~\cite{Unger:2015laa}  
\begin{widetext}
\begin{equation}
\frac{1}{\tau_{\rm int}^{A\gamma} (E)} = \frac{1}{\tau_b}
\left\{\begin{array}{ll}  \,
(E_b / E)^{\beta +1} & ~ E \leq E_b  \\
(1-\beta)/(1-\alpha) \left[\left( E_b/E \right)^{\alpha +1} -
  \left(E_b/E\right)^2 \right] +
\left(E_b/E\right)^2 & ~ E > E_b
\end{array} \right. \, ,
\label{tauint}
\end{equation}
\end{widetext}
where
\begin{equation}
\tau_b = \frac{ 2 \ E_b \ (1-\beta)} {c \, \pi \
  \sigma_{\rm res} \, A \, m_p \ \Gamma_{\rm res}
   \ n_0} \quad {\rm and} \quad
E_b = \frac{\varepsilon'_{\rm res} \ A \ m_p}{2 \varepsilon_0} ,
\end{equation}
and where the parameters characterizing the photodisintegration cross section are:
$\sigma_{\rm res} \approx 1.45\times 10^{-27}~{\rm cm}^2 \, A$,
$\Gamma_{\rm res} = 8~{\rm MeV}$, and $\varepsilon'_{\rm res} = 42.65
A^{-0.21} \, (0.925 A^{2.433})~{\rm MeV},$ for $A > 4$ ($A\leq
4$)~\cite{Karakula:1993he}. 

The photopion production process can also be
described by (\ref{tauint}). The parameters for the photopion
production cross section are: $\sigma_{\rm res} \simeq 5.0 \times
10^{-28}~{\rm cm}^2 \, A$, $\Gamma_{\rm res} = 150~{\rm MeV}$, and
$\varepsilon'_{\rm res} = (m_\Delta^2 - m_p^2)/(2 m_p) \simeq 340~{\rm
  MeV}$~\cite{ParticleDataGroup:2024cfk}. Photopion production
dominates over photodisintegration for nuclei with Lorentz factors
$\agt 10^{10}$.

Through examination of Fig.~\ref{fig:2} we can conclude that most UHECR nuclei would survive the trip (of about $5 \times 10^5~{\rm s}$) to the edge of Sombrero's nucleus without significant energy loss. It could be interesting to explore whether Sombrero satisfies all the characteristics of an UFA source~\cite{Unger:2015laa}. This would require showing that ``post-processing'' of UHECRs via photodisintegration in the bulge can naturally explain the entire spectrum and nuclear composition observed on Earth. As such an extensive study is beyond the scope of this paper, we defer this analysis to future work.

At this point a reality check is in
  order. To support our conjecture that the UHECR luminosity could be
  anti-correlated with the electromagnetic luminosity we must now identify
  which input parameters in the modeling of the Sombrero source lead to this
  anti-correlation. To establish UHECR acceleration up to above about
  $10^{11}~{\rm GeV}$ without nucleus photodisintegration in the
  acceleration region, we must simultaneously require a high potential
  drop and a low density of IR and FIR photons. The dynamo's emf is
  controlled by (\ref{V2}) and the average photon density by (\ref{n0}). We can see
  that the potential drop is directly proportional to $M_\bullet$ and $B$, whereas the average photon density is inversely proportional to the
  square of the Schwarzschild radius and therefore scales as
  $1/M_\bullet^2$. All in all, we could identify
  $M_\bullet$ as a key parameter in our model. Actually, we
  can further derive an estimate of the characteristic magnetic field strength $B$ by assuming pressure equilibrium between the magnetic field and the in-falling matter~\cite{Boldt:1999ge}
  \begin{equation}
    B \sim 10^4 \ \left(\frac{M_\bullet}{10^9 M_\odot}\right)^{-1}  \ \left(\frac{2.7 \ \dot M_{\rm
          Bondi}}{M_\odot/{\rm yr}} \right)^{1/2}~{\rm G} \, .
\label{B-field}
  \end{equation}    
Now, substituting (\ref{B-field}) into (\ref{V2}) it is straightforward to see that the
maximum energy of the cosmic rays is driven by the Bondi mass accretion
rate, see (\ref{V3}). The Bondi mass accretion rate is proportional to
$M_\bullet^2$~\cite{Bondi:1952ni} and the
density of (IR and FIR) photons in the acceleration region is suppressed by
$M_\bullet^2$. In summary, the emission of UHECRs is correlated with 
$M_\bullet^2$. As shown in Fig.~\ref{fig:2},  Sombrero's supermassive black hole, with
$M_\bullet \sim 10^9~M_\odot$, could allow for acceleration of
  UHECR nuclei without suffering catastrophic spallations in the acceleration region.

In addition, a re-interpretation of the results from an investigation presented
elsewhere~\cite{Moncada:2017hvq} substantiates our conjecture. On the one hand, the analyses of four supermassive black
  holes (living inside galaxies NGC 2768, NGC 3610, NGC 3613, and NGC 4125) with masses ${\cal O} (10^8
  M_\odot)$ seem to indicate that considerable nucleus photodisintegration in the
  vicinity of the event horizon would lead to low UHECR luminosity above about $10^{11}~{\rm GeV}$. On the other hand, the study of the supermassive black hole
  inside NGC 2832, with $M_\bullet \sim 10^{10} M_\odot$, suggests a
  low FIR photon density near the horizon, allowing UHECR nuclei to be
  accelerated up to the maximum observed energies on Earth.  In closing, a
  brief word of caution is necessary. The analyses of NGC 2768, NGC
  3610, NGC 3613, and NGC 4125 carried out in~\cite{Moncada:2017hvq} rely on
  several parameters whose estimates have large theoretical uncertainties. 
 
As an immediate spinoff of our conjecture, we could also argue that sources of UHECRs
are not sources of high energy neutrinos and vice versa. Note that
after UHECR nuclei escape unscathed from
the galaxy core, they could remain trapped in the source
environment by the average
large-scale magnetic field $\bar B_{\rm bulge}$. If this were the
case, some nuclei could undergo photodisintegration and subsequent
decay of the stripped neutrons would yield a flux of electron
antineutrinos. However, the total energy output in antineutrinos of
Sombrero results in a very low neutrino flux, which would be undetectable on
Earth. Actually, the diffuse antineutrino flux from all
Sombrero-type sources would also be out of the reach of next-generation
experiments, see
Fig.~5 in~\cite{Unger:2015laa}.

A point
  worth noting at this juncture is that this argument is supported by
  IceCube data. Firstly, the search for a possible correlation between
  the arrival direction of UHECRs and cosmic neutrinos returned only 
  null results~\cite{IceCube:2022osb}. Secondly, the astrophysical
  high-energy neutrino flux observed with IceCube may be as large as
  $10^{-7}~{\rm GeV \ cm^{-2} \ s^{-1} \ sr^{-1}}$ around 30~TeV~\cite{IceCube:2015gsk}. It
  was noted in~\cite{Murase:2015xka} that if sources of TeV-PeV
  neutrinos are transparent to $\gamma$-rays, a strong tension with the
  isotropic diffuse $\gamma$-ray background measured by the {\it
    Fermi} satellite~\cite{Fermi-LAT:2014ryh} becomes unavoidable -- independently of the
  neutrino production mechanism. This suggest that the sources of the
  astrophysical neutrinos observed by IceCube are hidden cosmic ray
  accelerators~\cite{Fang:2022trf}.
  Lastly, the IceCube Collaboration
  has reported evidence for neutrino emission from the nearby active
  galaxy NGC 1068~\cite{IceCube:2022der}. The redshift-corrected
  isotropic equivalent neutrino luminosity in the neutrino energy
  range from 1.5~TeV to 15~TeV estimated by the IceCube Collaboration,
  $L_\nu = (2.9 \pm 1.1 {\rm stat}) \times 10^{42} ~{\rm erg/s}$, is
  significantly higher than the isotropic equivalent $\gamma$-ray 
  luminosity observed by Fermi-LAT of $1.6 \times 10^{41}~{\rm erg/s}$ in the
  energy range between 100~MeV and 100~GeV~\cite{Fermi-LAT:2019pir}, and higher than the
  upper limits in the
  energy region where the neutrino signal was observed reported by
  the MAGIC collaboration~\cite{MAGIC:2019fvw}.  For this reason, NGC
  1068 is thought to be an astrophysical hidden source, where
  neutrinos are produced through $\pi^\pm$ decay, while the photons from $\pi^0$ decay are absorbed by interactions at the
  source~\cite{Murase:2019vdl,Inoue:2019yfs,Anchordoqui:2021vms}. Of
  particular interest here, NGC 1068 has also been identified as a
  plausible UHECR accelerator as it has a small contribution to the
  medium-scale anisotropy signal reported by the Pierre Auger
  Collaboration~\cite{PierreAuger:2018qvk}.  Thus, NGC 1068 provides a
  unique target to test our conjecture using future data from both
  IceCube and UHECR
  observatories.\footnote{Currently, IceCube detects an excess from the
  direction of NGC 1068 with a $4.2\sigma$ significance~\cite{IceCube:2022der}, which is
  still below discovery level.} However, information from a single identified source only
  provides limited insight into the general properties of the entire
  source population. For instance, the majority of high-energy astrophysical
  neutrinos detected by IceCube form a ``diffuse flux'' that appears
  to come from all directions in the sky, suggesting they originate
  from a vast population of individually faint sources that are too
  distant to resolve. Studying multiple sources and their collective
  properties would be essential for understanding whether, as
  conjectured herein, IceCube neutrinos originate in hidden sources, which are
  not UHECR accelerators. First steps in this direction have been taken in~\cite{Sommani:2024sbp,Abbasi:2025tas,Yang:2025lmb,Ma:2025tpg}.

We end with an observation: UHECRs could also be accelerated in Sombrero's
large-scale jet/lobes~\cite{He:2024vnm}. For a total jet power $\sim 10^{42}~{\rm TeV/s}$~\cite{Mezuca}, with
 intrinsic velocity $\sim 0.2 c$~\cite{Hada:2013tqa}, Fermi acceleration at shocks leads to a maximum cosmic ray energy
 \begin{equation}
E_{\rm max} \sim 10^{11.3} \left(\frac{Z}{26} \right) \ \left(\frac{\overline B}{5~\mu{\rm G}}\right) \left(\frac{v_{\rm sh}}{0.2 c}\right) \left(\frac{r_{\rm acc}}{10~kpc} \right)~{\rm GeV} \,
\end{equation}
where $\overline B$ is the average magnetic field strength at the acceleration site, $v_{\rm sh}$ is the velocity of the shock, and $r_{\rm acc}$ is the shock scale~\cite{He:2024vnm}.

\section{$\bm{\gamma}$-Ray Emission from Sombrero Galaxy}
\label{sec:5}

As a sharp reader might have noticed, the emitted spectrum of curvature
radiation peaks in the TeV energy band~\cite{Levinson:2000nx}. To be precise, most of the photons are emitted at a critical energy~\cite{Levinson:2002ea}  
\begin{equation}
 E_\gamma \sim 3 \times 10^3 \ (Z/A)^3 \ \dot M_{\rm Bondi}^{3/2}~{\rm
   TeV} \, .
\end{equation}
Now, these $\gamma$-rays would scatter (head-on) with the IR 
photons in the nucleus to produce electron-positron pairs. In this section, we
investigate whether these collisions could 
lead to a suppression of the curvature $\gamma$-ray flux. 

Two photons of energy $\varepsilon$ and $E_\gamma$ can produce an $e^+e^-$
pair in a collision if
\begin{equation}
E_\gamma \ \varepsilon \geq 2 \ m_e^2/ (1 - \cos \alpha) \,,
\end{equation}
where $m_e$ is the electron mass and $\alpha$ is the angle between the
incident directions of the photons. For a fixed $\gamma$-ray of energy $E_\gamma$, the pair production
cross-section $\sigma_{\gamma \gamma}$ depends on the IR photon
energy $\varepsilon$. The normalized center-of-mass energy squared of the
photons can be written as
\begin{equation}
  s = \frac{2 E_\gamma \ \varepsilon}{4m_e^2} \,,
\end{equation}
where we have assumed that $\alpha \sim \pi/2$. Starting from zero at the
threshold energy $\varepsilon_{\rm th}$, or equivalently $s = 1$, the cross-section rises steeply to a maximum
of $\sigma_T /4$ at $s \sim 2$ and falls off
as $s^{-1}$ for energies $\varepsilon \gg \varepsilon_{\rm th}$, where $\sigma_T$ is the Thomson cross section~\cite{Coppi:1990}.

We describe the absorption of the $\gamma$-rays by calculating the
dimensionless optical depth $\tau$. The emitted $\gamma$-ray intensity varies
as
\begin{equation}
I_\gamma \sim I_{\gamma_0} \ e^{-\tau} \,,
\end{equation}  
where $I_{\gamma_0}$ is the produced intensity via curvature
radiation. The optical depth characterizing the absorption of a $\gamma$-ray
with energy $E_\gamma$ propagating through an isotropic radiation field with
spectral distribution $n(\varepsilon)$ can be written as~\cite{Gould:1967zzb}
\begin{equation}
  \tau (E_\gamma) \sim r_{\rm nuc} \int_{2 m_e^2/E_\gamma}^\infty d
  \varepsilon  \ n(\varepsilon) \ \sigma_{\gamma \gamma} \, .
\label{opticaldepth}
\end{equation} 
Actually, the optical depth for a $\gamma$-ray of energy $E_\gamma$
is essentially determined by a relatively narrow band of photons
(centered at $s=2$). Following~\cite{Herterich:1974}, we approximate the cross section by a
rectangular function with height $\overline \sigma_{\gamma \gamma}
\sim \sigma_T/4$ and width $2.5 \varepsilon_{\rm th}$ to get from
(\ref{opticaldepth})
\begin{equation}
\tau (E_\gamma) \sim 2.5 \ \varepsilon_{\rm th} \ r_{\rm nuc} \ \overline \sigma_{\gamma
  \gamma} \  n(2\varepsilon_{\rm th}) \, .
\label{opa}
\end{equation}  
We remind the reader that our choice for the fiducial values of the $\alpha$ and $\beta$ parameters simplifies the analytic calculation in the region relevant  to nucleus photodisintegration. However, these fiducial values are not descriptive of the energy window $\varepsilon \sim 1~{\rm eV}$, which is relevant for production of $e^+e^-$ pairs.
The useful approximation of (\ref{opa}) written in terms of the IR luminosity also gives the correct order of
magnitude~\cite{Rieger:2011ch}
\begin{equation}
  \tau (E_\gamma) \sim  \left(\frac{\overline{L_{\rm IR}^{\rm
          nuc}}}{10^{6.7} L_\odot}\right) \ \left(\frac{r_{\rm nuc}}{50
        r_s}\right)^{-1} \ \left(\frac{E_\gamma}{{\rm TeV}}\right) \, .
\end{equation}
Since the nucleus' IR luminosity is not expected to be much smaller than its FIR luminosity, we conclude that the $\gamma$-ray flux would be partially reabsorbed inside the
galaxy.\footnote{We expect no neutrino signal accompanying the
  $\gamma$-ray emission for curvature radiation -- a purely
  electromagnetic process.} 

A note of caution: the preceding estimate is subject to model
assumptions/parameters which are not well-constrained; e.g.,
the dominance of energy losses due to emission of curvature photons is
contingent on UHECR experiencing large gaps, with $\xi \agt 0.1 r_s$ to reach $Ze V_{\rm
  eff} \agt 10^{11}~{\rm GeV}$. These systematic uncertainties lead to
an extremely broad range (several orders of magnitude) of possible
outcomes for the $\gamma$-ray
luminosity $L_\gamma$. A precise determination of $L_\gamma$ needs a step-by-step, computer-based numerical analysis,
which is beyond the scope of this paper. Needless to say, if an excess
of  $\gamma$-rays pointing back to Sombrero  were observed, then 
an investigation on the fraction of curvature photons escaping the galaxy 
would be important to conduct.

\section{Conclusions}
\label{sec:6}

Very recently, the back-tracked trajectories from two of the highest energy
cosmic rays (those with the smallest localization
 uncertainties on the source direction) lead to an unexpected result: none of these trajectories
points to {\it visible} prominent sources~\cite{Unger:2025xkk}. This
null result has been taken as evidence for UHECRs to originate in
transient sources. In this paper, we have explored an alternative
perspective: we conjectured that {\it accelerators of UHECRs anti-correlate
with the electromagnetic source power}. To support our conjecture we
investigated the feasibility of accelerating UHECRs in the
low-luminosity active galactic nucleus of the Sombrero galaxy, which
has been highlighted as a
distinctly possible candidate source~\cite{He:2024vnm}. We have shown
that UHECR nuclei
can be efficiently accelerated in the vicinity of the supermassive ($M_\bullet \sim 10^9 M_\odot$)
black hole 
living at the center of the galaxy.

UHECR acceleration proceeds via the Blandford-Znajek mechanism~\cite{Blandford:1977ds} by
extracting rotational energy from the spinning black hole, akin to a
unipolar dynamo. This process wires the black hole's magnetic field
through its rotation to generate powerful electromagnetic
outflows. Particles could then be accelerated to extreme energies, in
the vacuum gaps of the black hole's magnetosphere.

To establish UHECR acceleration up to above about
  $10^{11}~{\rm GeV}$ without nucleus photodisintegration in the
  acceleration region, we simultaneously required a high potential
  drop in the gap and a low density of IR and FIR photons in the
  vicinity of the event horizon. We showed that the
  dynamo's
  emf,  controlled by the potential drop, is directly proportional to the
  black hole mass $M_\bullet$,
  whereas the average photon density is inversely proportional to the
  square of the Schwarzschild radius and therefore scales as
  $1/M_\bullet^2$. Further, the potential drop is also proportional
  to the magnetic field strength, which scales as 
$1/M_\bullet$ and is proportional to the Bondi mass accretion
rate. Since the Bondi mass accretion rate is proportional
to $M_\bullet^2$ and the
density of (IR and FIR) photons in the acceleration region is suppressed by
$M_\bullet^2$, we reached the conclusion that the emission of UHECRs
is correlated with
the square of the black hole mass. A comparison with a study~\cite{Moncada:2017hvq} of galaxies
harboring supermassive
black holes with $M_\bullet \sim 10^8 M_\odot$ seems to indicate
that  Sombrero's supermassive black hole may be
near the critical mass required for emission of highest energy
cosmic rays. This is because in galaxies with black holes of
$M_\bullet \sim 10^8 M_\odot$ a considerable nucleus photodisintegration in the
  vicinity of the event horizon would lead to a low UHECR luminosity above about $10^{11}~{\rm GeV}$.

  An intrinsic property of UHECR sources satisfying our conjecture is
  that their neutrino emission is strongly suppressed. We have shown
  that current IceCube data~\cite{IceCube:2015gsk} fall in line with
  this prediction and that the most promising neutrino emitter, NGC
  1068, carries in itself substantial 
  evidence~\cite{IceCube:2022der}.
But of course a single data point is insufficient for characterizing
the entire source population. Future IceCube
  observations could help determine whether high-energy neutrinos come
  dominantly from hidden sources (potentially confirming the existence
  of sources that are not visible in other wavelengths), and could 
  concomintantly serve as an indirect test for the ideas discussed in this paper.

\section*{Acknowledgments}

We thank Michael Unger for some valuable discussion. The work of L.A.A. and K.P.C. is supported by the U.S. National Science
Foundation (NSF Grant PHY-2412679).


\begin{thebibliography}{99}




\bibitem{Waxman:1995dg}
E.~Waxman,
{\color{rossoCP3} Cosmological origin for cosmic rays above $10^{19}~{\rm eV}$},
Astrophys. J. Lett. \textbf{452}, L1-L4 (1995)
doi:10.1086/309715
[arXiv:astro-ph/9508037 [astro-ph]].

  
\bibitem{Gaisser:1997aw}
T.~K.~Gaisser,
{\color{rossoCP3}  Neutrino astronomy: Physics goals, detector parameters},
[arXiv:astro-ph/9707283 [astro-ph]].


\bibitem{Anchordoqui:2018qom}
L.~A.~Anchordoqui,
{\color{rossoCP3} Ultra-high-energy cosmic rays},
Phys. Rept. \textbf{801}, 1-93 (2019)
doi:10.1016/j.physrep.2019.01.002
[arXiv:1807.09645 [astro-ph.HE]].

\bibitem{Unger:2025xkk}
M.~Unger and G.~R.~Farrar,
{\color{rossoCP3} The Galactic magnetic field and UHECR deflections},
[arXiv:2502.15876 [astro-ph.HE]].

\bibitem{Unger:2023hnu}
M.~Unger and G.~R.~Farrar,
 {\color{rossoCP3} Where did the Amaterasu particle come from?},
Astrophys. J. Lett. \textbf{962}, no.1, L5 (2024)
doi:10.3847/2041-8213/ad1ced
[arXiv:2312.13273 [astro-ph.HE]].

\bibitem{Boldt:1999ge}
E.~Boldt and P.~Ghosh,
 {\color{rossoCP3} Cosmic rays from remnants of quasars?},
Mon. Not. Roy. Astron. Soc. \textbf{307}, 491-494 (1999)
doi:10.1046/j.1365-8711.1999.02600.x
[arXiv:astro-ph/9902342 [astro-ph]].

\bibitem{Blandford:1977ds}
R.~D.~Blandford and R.~L.~Znajek,
 {\color{rossoCP3} Electromagnetic extractions of energy from Kerr black holes},
Mon. Not. Roy. Astron. Soc. \textbf{179}, 433-456 (1977)
doi:10.1093/mnras/179.3.433


\bibitem{Unger:2015laa}
M.~Unger, G.~R.~Farrar and L.~A.~Anchordoqui,
 {\color{rossoCP3}  Origin of the ankle in the ultrahigh energy cosmic ray spectrum, and of the extragalactic protons below it},
Phys. Rev. D \textbf{92}, no.12, 123001 (2015)
doi:10.1103/PhysRevD.92.123001
[arXiv:1505.02153 [astro-ph.HE]].


\bibitem{He:2024vnm}
H.~N.~He, E.~Kido, K.~K.~Duan, Y.~Yang, R.~Higuchi, Y.~Z.~Fan, T.~Wang, L.~Y.~Jiang, R.~L.~Li and B.~Y.~Zhu, \textit{et al.}
 {\color{rossoCP3}  Evidence for the Sombrero galaxy as an accelerator of the highest-energy cosmic rays},
[arXiv:2412.11966 [astro-ph.HE]].

\bibitem{Goold} K. Goold {\it et al.},  {\color{rossoCP3}  ReveaLLAGN
    0: First look at JWST MIRI data of Sombrero and NGC 1052},
  Astrophys. J. {\bf 966}, 204 (2024)
doi:10.3847/1538-4357/ad3065
[arXiv:2307.01252 [astro-ph.GA]]  



\bibitem{Kormendy:1988}
  J.~Kormendy,
 {\color{rossoCP3} Evidence for a central dark mass in NGC 4594 (the
   Sombrero galaxy)},
 Astrophys. J. {\bf 335}, 40-56 (1988)
doi:10.1086/166904 

\bibitem{Kormendy:1996}
J. Kormendy, R. Bender, E. A. Ajhar, A. Dressler, S. M. Faber,
K. Gebhardt, C. Grillmair, T. R. Lauer, D. Richstone, and S. Tremaine,
 {\color{rossoCP3} Hubble Space Telescope spectroscopic evidence for a
   $1 \times 10^9 M_\odot$ black hole in NGC 4594}
 Astrophys. J. {\bf 473}, L91-L94 (1996)
doi:10.1086/310399 

\bibitem{Jardel:2011dh}
J.~R.~Jardel, K.~Gebhardt, J.~Shen, D.~Fisher, J.~Kormendy, J.~Kinzler, T.~R.~Lauer, D.~Richstone and K.~Gultekin,
 {\color{rossoCP3} Orbit-Based Dynamical Models of the Sombrero Galaxy (NGC 4594)},
Astrophys. J. \textbf{739}, 21 (2011)
doi:10.1088/0004-637X/739/1/21
[arXiv:1107.1238 [astro-ph.CO]].


\bibitem{Burbidge:1957vc}
M.~E.~Burbidge, G.~R.~Burbidge, W.~A.~Fowler and F.~Hoyle,
 {\color{rossoCP3} Synthesis of the elements in stars},
Rev. Mod. Phys. \textbf{29}, 547-650 (1957)
doi:10.1103/RevModPhys.29.547




\bibitem{Anchordoqui:1999aj}
L.~A.~Anchordoqui, M.~T.~Dova, T.~P.~McCauley, S.~Reucroft and J.~D.~Swain,
 {\color{rossoCP3} Possible explanation for the tail of the cosmic ray spectrum},
Phys. Lett. B \textbf{482}, 343-348 (2000)
doi:10.1016/S0370-2693(00)00534-7
[arXiv:astro-ph/9912081 [astro-ph]].


\bibitem{Anchordoqui:2000sk}
L.~A.~Anchordoqui, M.~T.~Dova, T.~P.~McCauley, T.~C.~Paul, S.~Reucroft and J.~D.~Swain,
 {\color{rossoCP3} A Pot of gold at the end of the cosmic ``raynbow''?},
Nucl. Phys. B Proc. Suppl. \textbf{97}, 203-206 (2001)
doi:10.1016/S0920-5632(01)01264-6
[arXiv:astro-ph/0006071 [astro-ph]].


\bibitem{Farrar:2024zsm}
G.~R.~Farrar,
{\color{rossoCP3} Binary neutron star mergers as the source of the highest energy cosmic rays},
Phys. Rev. Lett. \textbf{134}, no.8, 081003 (2025)
doi:10.1103/PhysRevLett.134.081003
[arXiv:2405.12004 [astro-ph.HE]].

\bibitem{Zhang:2024sjp}
B.~T.~Zhang, K.~Murase, N.~Ekanger, M.~Bhattacharya and S.~Horiuchi,
{\color{rossoCP3}  Ultraheavy ultrahigh-energy cosmic rays},
[arXiv:2405.17409 [astro-ph.HE]].



\bibitem{Jansson:2012pc}
R.~Jansson and G.~R.~Farrar,
{\color{rossoCP3} A new model of the Galactic magnetic field},
Astrophys. J. \textbf{757}, 14 (2012)
doi:10.1088/0004-637X/757/1/14
[arXiv:1204.3662 [astro-ph.GA]].

\bibitem{Unger:2023lob}
M.~Unger and G.~R.~Farrar,
{\color{rossoCP3} The coherent magnetic field of the Milky Way},
Astrophys. J. \textbf{970}, no.1, 95 (2024)
doi:10.3847/1538-4357/ad4a54
[arXiv:2311.12120 [astro-ph.GA]].

\bibitem{Korochkin:2024yit}
A.~Korochkin, D.~Semikoz and P.~Tinyakov,
{\color{rossoCP3}  The coherent magnetic field of the Milky Way halo, the Local Bubble, and the Fan region},
Astron. Astrophys. \textbf{693}, A284 (2025)
doi:10.1051/0004-6361/202451440
[arXiv:2407.02148 [astro-ph.GA]].

\bibitem{Korochkin:2025ugg}
A.~Korochkin, D.~Semikoz and P.~Tinyakov,
{\color{rossoCP3}  UHECR deflections in the Galactic magnetic field},
[arXiv:2501.16158 [astro-ph.HE]].

\bibitem{PierreAuger:2022axr}
P.~Abreu \textit{et al.} [Pierre Auger],
{\color{rossoCP3} Arrival directions of cosmic rays above 32~EeV from Phase one of the Pierre Auger Observatory},
Astrophys. J. \textbf{935}, no.2, 170 (2022)
doi:10.3847/1538-4357/ac7d4e
[arXiv:2206.13492 [astro-ph.HE]].




\bibitem{McQuinn}
K. B. W. McQuinn, E. D. Skillman, A. E. Dolphin, D. Berg, and R. Kennicutt,
{\color{rossoCP3} The distance to M104}
Astron. J. {\bf 152}, 144 (2016)
doi:10.3847/0004-6256/152/5/144


\bibitem{Tempel:2006qu}
E.~Tempel and P.~Tenjes,
{\color{rossoCP3} Line-of-sight velocity dispersions and a mass distribution model of the Sa galaxy NGC 4594},
Mon. Not. Roy. Astron. Soc. \textbf{371}, 1269-1279 (2006)
doi:10.1111/j.1365-2966.2006.10741.x
[arXiv:astro-ph/0606680 [astro-ph]].

\bibitem{Bondi:1952ni}
H.~Bondi,
 {\color{rossoCP3} On spherically symmetrical accretion},
Mon. Not. Roy. Astron. Soc. \textbf{112}, 195 (1952)
doi:10.1093/mnras/112.2.195

\bibitem{Pellegrini:2003pu}
S.~Pellegrini, A.~Baldi, G.~Fabbiano and D.~W.~Kim,
{\color{rossoCP3} An XMM-newton and chandra investigation of the
  nuclear accretion in the sombrero galaxy (NGC 4594),}
Astrophys. J. \textbf{597}, 175-185 (2003)
doi:10.1086/378235
[arXiv:astro-ph/0307142 [astro-ph]].

\bibitem{Li:2010uk}
Z.~Li, C.~Jones, W.~R.~Forman, R.~P.~Kraft, D.~V.~Lal, R.~Di Stefano, L.~R.~Spitler, S.~Tang, Q.~D.~Wang and M.~Gilfanov, \textit{et al.}
{\color{rossoCP3} X-ray emission from the Sombrero galaxy: a galactic-scale outflow},
Astrophys. J. \textbf{730}, 84 (2011)
doi:10.1088/0004-637X/730/2/84
[arXiv:1009.5767 [astro-ph.CO]].


\bibitem{Gallimore:2006dp}
J.~F.~Gallimore, D.~J.~Axon, C.~P.~O'Dea, S.~A.~Baum and A.~Pedlar,
{\color{rossoCP3} A survey of kiloparsec-scale radio outflows in radio-quiet active galactic nuclei},
Astron. J. \textbf{132}, 546-569 (2006)
doi:10.1086/504593
[arXiv:astro-ph/0604219 [astro-ph]].

\bibitem{Hada:2013tqa}
K.~Hada, A.~Doi, H.~Nagai, M.~Inoue, M.~Honma, M.~Giroletti and G.~Giovannini,
{\color{rossoCP3}  Evidence for nuclear radio jet and its structure
  down to $< 100$ Schwarzschild radii in the center of the Sombrero galaxy (M 104, NGC 4594)},
Astrophys. J. \textbf{779}, 6 (2013)
doi:10.1088/0004-637X/779/1/6
[arXiv:1310.0488 [astro-ph.CO]].

\bibitem{Yang}
  Y. Yang, J.-T.  Li, T.  Wiegert, Z.  Li, F.  Guo, J. Irwin, Q. D.
  Wang, R.-J. Dettmar, R.  Beck, J. English, and L.  Ji,
 {\color{rossoCP3}  CHANG-ES. XXX. 10~kpc radio lobes in the Sombrero galaxy},
Astrophys. J. \textbf{966}, 213 (2024)
doi:10.3847/1538-4357/ad37fd
[arXiv:2403.16682 [astro-ph.GA]]


\bibitem{Krause:2005bt}
M.~Krause, R.~Wielebinski and M.~Dumke,
{\color{rossoCP3}  Radio polarization and sub-mm observations of the sombrero galaxy (NGC 4594): large-scale magnetic field configuration and dust emission},
Astron. Astrophys. \textbf{448}, 133 (2006)
doi:10.1051/0004-6361:20053789
[arXiv:astro-ph/0510796 [astro-ph]].



\bibitem{Cohen:2020}
R. E. Cohen, P. Goudfrooij, M. Correnti, O. Y. Gnedin, W. E. Harris, R. Chandar, T. H. Puzia, R. Sanchez-Janssen, 
{\color{rossoCP3} The strikingly metal-rich halo of the Sombrero galaxy},
Astrophys. J. {\bf 890}, 52 (2020)
doi:10.3847/1538-4357/ab64e9
[arXiv:2001.01670 [astro-ph.GA]]


\bibitem{Young:1989}
  J. S. Young, S.  Xie, J. D. P  Kenney,  W. L. Rice,
 {\color{rossoCP3} Global properties of infrared bright galaxies}
Astrophys. J. Supp. {\bf 70}, 699-722 (1989)
doi:10.1086/191355

\bibitem{IRAS}
Joint IRAS Science Working Group, 
{\color{rossoCP3} IRAS Point Source Catalog}, (Washington, DC: GPO, 1985).

\bibitem{Sutter:2022}
  J. Sutter and D. Fadda,
{\color{rossoCP3} A molecular gas ring hidden in the Sombrero galaxy},  
Astrophys. J. {\bf 941}, 47 (2022)
doi:10.3847/1538-4357/ac9d8f
[arXiv:2210.13527 [astro-ph.GA]]

\bibitem{Bendo:2006pc}
G.~J.~Bendo, B.~A.~Buckalew, D.~A.~Dale, B.~T.~Draine, R.~D.~Joseph, R.~C.~Kennicutt, K.~Sheth, J.~D.~T.~Smith, F.~Walter and D.~Calzetti, \textit{et al.}
{\color{rossoCP3}  Spitzer and JCMT Observations of the active galactic nucleus in the Sombrero galaxy (NGC 4594)},
Astrophys. J. \textbf{645}, 134-147 (2006)
doi:10.1086/504033
[arXiv:astro-ph/0603160 [astro-ph]].












\bibitem{Penrose:1969pc} 
  R.~Penrose,
    {\color{rossoCP3} Gravitational collapse: The role of general relativity},
  Riv.\ Nuovo Cim.\  {\bf 1}, 252 (1969)
  [Gen.\ Rel.\ Grav.\  {\bf 34}, 1141 (2002)].

\bibitem{Penrose:1971uk} 
  R.~Penrose and R.~M.~Floyd,
  {\color{rossoCP3}   Extraction of rotational energy from a black hole},
  Nature {\bf 229}, 177 (1971).

\bibitem{Christodoulou:1970wf}
D.~Christodoulou,
  {\color{rossoCP3}  Reversible and irreversible transforations in black hole physics},
Phys. Rev. Lett. \textbf{25}, 1596-1597 (1970)
doi:10.1103/PhysRevLett.25.1596

\bibitem{Znajek}
R.~L.~Znajek,
 {\color{rossoCP3} The electric and magnetic conductivity of a Kerr hole},
Mon. Not. Roy. Astron. Soc. \textbf{185}, 833-840 (1978)
doi:10.1093/mnras/185.4.833




\bibitem{Krolik:1999yc}
J.~H.~Krolik,
 {\color{rossoCP3}  Magnetized accretion inside the marginally stable orbit around a black hole},
Astrophys. J. Lett. \textbf{515}, L73-L76 (1999)
doi:10.1086/311979
[arXiv:astro-ph/9902267 [astro-ph]].

\bibitem{DiMatteo:1998bf}
T.~Di Matteo, A.~C.~Fabian, M.~J.~Rees, C.~L.~Carilli and R.~J.~Ivison,
 {\color{rossoCP3}  Strong observational constraints on advection-dominated accretion in the cores of elliptical galaxies},
Mon. Not. Roy. Astron. Soc. \textbf{305}, 492 (1999)
doi:10.1046/j.1365-8711.1999.02334.x
[arXiv:astro-ph/9807245 [astro-ph]].
  

\bibitem{Boldt:2000dx}
E.~Boldt and M.~Loewenstein,
 {\color{rossoCP3}  Cosmic ray generation by quasar remnants: Constraints and implications},
Mon. Not. Roy. Astron. Soc. \textbf{316}, L29 (2000)
doi:10.1046/j.1365-8711.2000.03768.x
[arXiv:astro-ph/0006221 [astro-ph]].

\bibitem{Narayan:1994is}
R.~Narayan and I.~Yi,
 {\color{rossoCP3}  Advection dominated accretion: Underfed black holes and neutron stars},
Astrophys. J. \textbf{452}, 710 (1995)
doi:10.1086/176343
[arXiv:astro-ph/9411059 [astro-ph]].

\bibitem{Levinson:2010fc}
A.~Levinson and F.~Rieger, 
 {\color{rossoCP3}  Variable TeV emission as a manifestation of jet formation in M87?},
Astrophys. J. \textbf{730}, 123 (2011)
doi:10.1088/0004-637X/730/2/123
[arXiv:1011.5319 [astro-ph.HE]].

\bibitem{Neronov:2009zz}
A.~Y.~Neronov, D.~V.~Semikoz and I.~I.~Tkachev,
 {\color{rossoCP3}  Ultra-high energy cosmic ray production in the polar cap regions of black hole magnetospheres},
New J. Phys. \textbf{11}, 065015 (2009)
doi:10.1088/1367-2630/11/6/065015
[arXiv:0712.1737 [astro-ph]].


\bibitem{Katsoulakos:2017byr}
G.~Katsoulakos and F.~M.~Rieger,
  {\color{rossoCP3}   Magnetospheric gamma-ray emission in active galactic nuclei},
Astrophys. J. \textbf{852}, no.2, 112 (2018)
doi:10.3847/1538-4357/aaa003
[arXiv:1712.04203 [astro-ph.HE]].



\bibitem{Rieger:2011ch} 
  F.~M.~Rieger,
    {\color{rossoCP3} Non-thermal processes in black-hole-jet magnetospheres},
  Int.\ J.\ Mod.\ Phys.\ D {\bf 20}, 1547 (2011)
  doi:10.1142/S0218271811019712
  [arXiv:1107.2119 [astro-ph.CO]].

\bibitem{Ochelkov} 
Y. P. Ochelkov and V. V. Usov,
{\color{rossoCP3} Curvature radiation of relativistic particles in the
  magnetosphere of pulsars}
 Astrophys.\ Space Sci.\  {\bf 69},  439 (1980)
doi:10.1007/BF00661929.

  
\bibitem{Moncada:2017hvq}
R.~J.~Moncada, R.~A.~Colon, J.~J.~Guerra, M.~J.~O'Dowd and L.~A.~Anchordoqui,
  {\color{rossoCP3} Ultrahigh energy cosmic ray nuclei from remnants of dead quasars},
JHEAp \textbf{13-14}, 32-45 (2017)
doi:10.1016/j.jheap.2017.04.001
[arXiv:1702.00053 [astro-ph.HE]].





\bibitem{Stecker:1969fw} 
  F.~W.~Stecker,
 {\color{rossoCP3}  Photodisintegration of ultrahigh-energy cosmic rays by the universal radiation field},
  Phys.\ Rev.\  {\bf 180}, 1264 (1969).
  doi:10.1103/PhysRev.180.1264


\bibitem{Anchordoqui:2006pe} 
  L.~A.~Anchordoqui, J.~F.~Beacom, H.~Goldberg, S.~Palomares-Ruiz and T.~J.~Weiler,
  {\color{rossoCP3} TeV $\gamma$-rays and neutrinos from photodisintegration of nuclei in Cygnus OB2},
  Phys.\ Rev.\ D {\bf 75}, 063001 (2007)
  doi:10.1103/PhysRevD.75.063001
  [astro-ph/0611581].


\bibitem{Karakula:1993he} 
  S.~Karakula and W.~Tkaczyk,
  {\color{rossoCP3}  The formation of the cosmic ray energy spectrum by a photon field},
  Astropart.\ Phys.\  {\bf 1}, 229 (1993).
  doi:10.1016/0927-6505(93)90023-7


\bibitem{ParticleDataGroup:2024cfk}
S.~Navas \textit{et al.} [Particle Data Group],
 {\color{rossoCP3} Review of particle physics},
Phys. Rev. D \textbf{110}, no.3, 030001 (2024)
doi:10.1103/PhysRevD.110.030001

\bibitem{IceCube:2022osb}
A.~Albert \textit{et al.} [IceCube, Pierre Auger, Telescope Array, Auger and ANTARES],
 {\color{rossoCP3} Search for spatial correlations of neutrinos with ultra-high-energy cosmic rays},
Astrophys. J. \textbf{934}, no.2, 164 (2022)
doi:10.3847/1538-4357/ac6def
[arXiv:2201.07313 [astro-ph.HE]].

\bibitem{IceCube:2015gsk}
M.~G.~Aartsen \textit{et al.} [IceCube],
 {\color{rossoCP3}  A combined maximum-likelihood analysis of the high-energy astrophysical neutrino flux measured with IceCube},
Astrophys. J. \textbf{809}, no.1, 98 (2015)
doi:10.1088/0004-637X/809/1/98
[arXiv:1507.03991 [astro-ph.HE]].



\bibitem{Murase:2015xka}
K.~Murase, D.~Guetta and M.~Ahlers,
 {\color{rossoCP3} Hidden cosmic-ray accelerators as an origin of TeV-PeV cosmic neutrinos},
Phys. Rev. Lett. \textbf{116}, no.7, 071101 (2016)
doi:10.1103/PhysRevLett.116.071101
[arXiv:1509.00805 [astro-ph.HE]].


\bibitem{Fermi-LAT:2014ryh}
M.~Ackermann \textit{et al.} [Fermi-LAT],
 {\color{rossoCP3} The spectrum of isotropic diffuse gamma-ray emission between 100 MeV and 820 GeV},
Astrophys. J. \textbf{799}, 86 (2015)
doi:10.1088/0004-637X/799/1/86
[arXiv:1410.3696 [astro-ph.HE]].

\bibitem{Fang:2022trf}
K.~Fang, J.~S.~Gallagher and F.~Halzen,
 {\color{rossoCP3}  The TeV diffuse cosmic neutrino spectrum and the nature of astrophysical neutrino sources},
Astrophys. J. \textbf{933}, no.2, 190 (2022)
doi:10.3847/1538-4357/ac7649
[arXiv:2205.03740 [astro-ph.HE]].

\bibitem{IceCube:2022der}
R.~Abbasi \textit{et al.} [IceCube],
 {\color{rossoCP3}  Evidence for neutrino emission from the nearby active galaxy NGC 1068},
Science \textbf{378}, no.6619, 538-543 (2022)
doi:10.1126/science.abg3395
[arXiv:2211.09972 [astro-ph.HE]].

\bibitem{Fermi-LAT:2019pir}
M.~Ajello \textit{et al.} [Fermi-LAT],
 {\color{rossoCP3} The fourth Catalog of Active Galactic Nuclei Detected by the Fermi Large Area Telescope},
Astrophys. J. \textbf{892}, 105 (2020)
doi:10.3847/1538-4357/ab791e
[arXiv:1905.10771 [astro-ph.HE]].

\bibitem{MAGIC:2019fvw}
V.~A.~Acciari \textit{et al.} [MAGIC],
 {\color{rossoCP3}  Constraints on gamma-ray and neutrino emission from NGC 1068 with the MAGIC telescopes},
Astrophys. J. \textbf{883}, 135 (2019)
doi:10.3847/1538-4357/ab3a51
[arXiv:1906.10954 [astro-ph.HE]].

\bibitem{Murase:2019vdl}
K.~Murase, S.~S.~Kimura and P.~Meszaros,
 {\color{rossoCP3}  Hidden cores of active galactic nuclei as the origin of medium-energy neutrinos: Critical tests with the MeV gamma-ray connection},
Phys. Rev. Lett. \textbf{125}, no.1, 011101 (2020)
doi:10.1103/PhysRevLett.125.011101
[arXiv:1904.04226 [astro-ph.HE]].



\bibitem{Inoue:2019yfs}
Y.~Inoue, D.~Khangulyan and A.~Doi,
 {\color{rossoCP3}  On the origin of high-energy neutrinos from NGC 1068: The role of nonthermal coronal activity},
Astrophys. J. Lett. \textbf{891}, no.2, L33 (2020)
doi:10.3847/2041-8213/ab7661
[arXiv:1909.02239 [astro-ph.HE]].



\bibitem{Anchordoqui:2021vms}
L.~A.~Anchordoqui, J.~F.~Krizmanic and F.~W.~Stecker,
 {\color{rossoCP3}   High-energy neutrinos from NGC 1068},
PoS \textbf{ICRC2021}, 993 (2021)
doi:10.22323/1.395.0993
[arXiv:2102.12409 [astro-ph.HE]].

\bibitem{PierreAuger:2018qvk}
  A.~Aab \textit{et al.} [Pierre Auger],
  {\color{rossoCP3}   An indication of anisotropy in arrival directions of ultra-high-energy cosmic rays through comparison to the flux pattern of extragalactic gamma-ray sources},
Astrophys. J. Lett. \textbf{853}, no.2, L29 (2018)
doi:10.3847/2041-8213/aaa66d
[arXiv:1801.06160 [astro-ph.HE]].

\bibitem{Sommani:2024sbp}
G.~Sommani, A.~Franckowiak, M.~Lincetto and R.~J.~Dettmar,
 {\color{rossoCP3}  Two 100 TeV neutrinos coincident with the Seyfert galaxy NGC 7469},
Astrophys. J. \textbf{981}, no.2, 103 (2025)
doi:10.3847/1538-4357/adb031
[arXiv:2403.03752 [astro-ph.HE]].

  
\bibitem{Abbasi:2025tas}
R.~Abbasi  \textit{et al.} [IceCube],
 {\color{rossoCP3}  Evidence for neutrino emission from X-ray bright active galactic nuclei with IceCube},
[arXiv:2510.13403 [astro-ph.HE]].


\bibitem{Yang:2025lmb}
Q.~R.~Yang, X.~B.~Chen, R.~Y.~Liu, X.~Y.~Wang and M.~Lemoine,
 {\color{rossoCP3}  On the origin of {\textasciitilde} 100~TeV neutrinos from the Seyfert galaxy NGC 7469},
[arXiv:2510.19662 [astro-ph.HE]].


\bibitem{Ma:2025tpg}
Z.~P.~Ma, K.~Wang, Y.~Y.~Zuo, Y.~C.~Zou and Y.~H.~Huang,
 {\color{rossoCP3}  Seyfert galaxies as neutrino sources: An outflow$-$cloud interaction perspective},
[arXiv:2511.06707 [astro-ph.HE]].
  



\bibitem{Mezuca}
M. Mezuca and M. A. Prieto
 {\color{rossoCP3} Evidence of parsec-scale jets in low-luminosity active galactic nuclei},
Astrophys. J. {\bf 787}, 62 (2014)
doi:10.1088/0004-637X/787/1/62
[arXiv:1403.6675 [astro-ph.Ga]]
 

\bibitem{Levinson:2000nx} 
  A.~Levinson,
  {\color{rossoCP3} Particle acceleration and curvature TeV emission by rotating supermassive black holes},
  Phys.\ Rev.\ Lett.\  {\bf 85}, 912 (2000).
  doi:10.1103/PhysRevLett.85.912
  [arXiv:hep-ph/0002020 [hep-ph]].

 \bibitem{Levinson:2002ea}
A.~Levinson and E.~Boldt,
 {\color{rossoCP3} UHECR production by a compact black hole dynamo: Application to Sgr A*},
Astropart. Phys. \textbf{16}, 265-270 (2002)
doi:10.1016/S0927-6505(01)00116-5
[arXiv:astro-ph/0012314 [astro-ph]].

\bibitem{Coppi:1990}
P. S. Coppi and R. D. Blandford,
 {\color{rossoCP3} Reaction rates and energy distributions for
   elementary processes in relativistic pair plasmas},
Mon. Not. Roy. Astron. Soc. \textbf{245}, 453-469 (1990)
doi:10.1093/mnras/245.3.453

 

\bibitem{Gould:1967zzb}
R.~J.~Gould and G.~P.~Schreder,
 {\color{rossoCP3} Pair production in photon-photon collisions},
Phys. Rev. \textbf{155}, 1404-1407 (1967)
doi:10.1103/PhysRev.155.1404

\bibitem{Herterich:1974}
K. Herterich,
 {\color{rossoCP3} Absorption of gamma rays in intense $X$-ray sources},
 Nature {\bf 250}, 311 (1974)
 doi:10.1038/250311a0

\end{thebibliography}
\end{document}